\documentclass{article}

\usepackage{arxiv}

\usepackage[utf8]{inputenc} 
\usepackage[T1]{fontenc}    
\usepackage{hyperref}       
\usepackage{url}            
\usepackage{booktabs}       
\usepackage{amsfonts}       
\usepackage{nicefrac}       
\usepackage{microtype}      
\usepackage{lipsum}
\usepackage{graphicx}
\graphicspath{ {./images/} }
\usepackage{natbib}  

\title{Write on Paper, Wrong in Practice: Why LLMs Still Struggle with Writing Clinical Notes}

\author{
Kristina L. Kupferschmidt \\
School of Mathematical and Computational Sciences\\
University of Prince Edward Island\\
Canada \\
\texttt{kkupferschmidt@upei.ca}
\And
Kieran O'Doherty \\
Department of Psychology\\
University of Guelph\\
Canada \\
\texttt{ohdertk@uoguelph.ca}
\And
Joshua A. Skorburg \\
Department of Philosophy\\
University of Guelph\\
Canada \\
\texttt{skorburg@uoguelph.ca}
}

\begin{document}
\maketitle
\begin{center}
    \textbf{This paper has been accepted to AIES 2025.}
\end{center}
\vspace{1em}

\begin{abstract}
Large Language Models (LLMs) are often proposed as tools to streamline clinical documentation, a task viewed as both high-volume and low-risk. However, even seemingly straightforward applications of LLMs raise complex sociotechnical considerations to translate into practice. This case study, conducted at KidsAbility, a pediatric rehabilitation facility in Ontario, Canada examined the use of LLMs to support occupational therapists in reducing documentation burden.We conducted a qualitative study involving 20 clinicians who participated in pilot programs using two AI technologies: a general-purpose proprietary LLM and a bespoke model fine-tuned on proprietary historical documentation.

Our findings reveal that documentation challenges are sociotechnical in nature, shaped by clinical workflows, organizational policies, and system constraints. Four key themes emerged: (1) the heterogeneity of workflows, (2) the documentation burden is systemic and not directly linked to the creation of any single type of documentation, (3) the need for flexible tools and clinician autonomy, and (4) effective implementation requires mutual learning between clinicians and AI systems.

While LLMs show promise in easing documentation tasks, their success will depend on flexible, adaptive integration that supports clinician autonomy. Beyond technical performance, sustained adoption will require training programs and implementation strategies that reflect the complexity of clinical environments.
\end{abstract}


\section{Introduction}

Large Language Models (LLMs) have been heralded as a turning point in healthcare. Popular tools like GPT-4 have demonstrated performance on par with or exceeding physicians on benchmark evaluations such as the United States Medical Licensing Examination (USMLE) \citep{Brin2023-iz}, and are already being deployed in applications ranging from patient triage to diagnostic support. The speed of adoption has been unprecedented, with healthcare systems, vendors, and startups racing to integrate LLMs into clinical workflows \citep{Ma2024-oc, Peterson-Health-Technology-Institute-AI-Taskforce2025-aq}. This is coming to fruition in many ways, with health providers exploring LLMs as clinical assistants, scribes, and documentation tools. As adoption accelerates, these tools are increasingly positioned as transformative for clinical practice. Yet amid bold predictions about replacing diagnosticians, a more grounded question remains: can these models reliably handle the “easy” tasks? And if they can, will that performance hold up under the complexity of real-world settings?

Clinical documentation, especially when written in frequently used standard formats like SOAP (Subjective, Objective, Assessment, Plan) have been pitched as a low-risk and high-reward use case. SOAP notes are central to both care delivery and legal documentation for many health fields but are often criticized for their repetitive and time-consuming nature. Because SOAP notes follow a widely taught structure and account for the bulk of therapists’ record-keeping time, they’re often cast as “low-hanging fruit” for automation \citep{Biswas2024-kg}. 

However, our study challenges that assumption. We present a detailed qualitative case study of a pilot program testing LLMs in a pediatric rehabilitation clinic. The goal was to reduce clinician documentation time by helping occupational therapists generate SOAP notes more efficiently. In practice, clinicians begin with “scratch notes” (i.e. bullet points, memory cues, or brief sentences) that are written during the patient interaction and transform them into the SOAP format after the fact \citep{Amenyo2025-ku}. On paper, this task maps neatly onto current LLM capabilities: convert unstructured text into a consistent, templated output. Yet, deployment and follow-up interviews revealed layers of friction such as formatting mismatches, user distrust, workflow misalignment, and perceived threats to clinician autonomy. These issues stem not from poor language generation, but from deeper, systemic mismatches between real-world documentation practices and user expectations of model behaviours.

Prior research has explored both the potential and limitations of using LLMs for documentation. Some studies have found success in using LLMs to summarize visit notes, translate clinical jargon for patients, or assist with discharge summaries \citep{Ma2024-oc}. Yet even in these use cases, concerns persist about output quality, interpretability, alignment with clinician expectations, and the ability of AI systems to enhance, rather than disrupt, existing care workflows. 

Our findings suggest that while LLMs may be technically capable of generating documentation, their real-world effectiveness depends on a web of sociotechnical factors: the variability of clinician workflows, the adaptability of AI interfaces, the depth of user training, and the policies governing documentation practices. In other words, even structurally simple tasks require careful, context-aware implementation. \citet{Mollick2025-iu} promotes a similar sentiment, that widespread enthusiasm for LLMs may not yield the broader organizational benefits it promises. In particular, he emphasizes that in order to realize meaningful gains from tools like LLMs, organizations need to treat adoption as a process of organizational learning rather than focusing on the technical roll out. 

While much of the discourse around AI ethics in healthcare has focused on fairness, transparency, and data provenance and privacy, there is a tendency to treat technical feasibility as a foregone conclusion \citep{Satheakeerthy2025-pg, Ong2024-vx}. Specifically, in the context of building trust in the end-user and what a “trustworthy” system really consists of \citep{Burger2024-cj}. But, if even seemingly simple tasks like documentation remain difficult to implement effectively, then efficacy itself must be treated as an ethical concern. Deploying systems before they've demonstrated consistent utility in real clinical environments risks introducing not just inefficiency, but also harm through wasting clinician time, undermining trust, or producing documentation that is misaligned with standards of care.

\noindent The objective of this study was to assess the feasibility of using modern LLMs to reduce documentation burden in pediatric occupational therapy. Although our analysis is grounded in this specialized setting, the lessons learned are broadly applicable across healthcare domains. This case study makes three core contributions:
\begin{enumerate}
    \item It exposes the gap between LLM capabilities and their real-world clinical utility in documentation-heavy workflows.
    \item It demonstrates how deployment failures often stem not from model performance alone, but from misalignments between tool design and the clinical context.
    \item It underscores that healthcare documentation remains an unsolved challenge, even for seemingly straightforward tasks.
\end{enumerate}

Our findings offer a revealing test case for evaluating LLM readiness in healthcare. If we cannot successfully automate routine documentation, it raises critical concerns about deploying these models for more complex, high-stakes clinical applications.

\section{Background}
\subsection{Adoption of LLMs in Healthcare}
The integration of LLMs into healthcare has accelerated rapidly over the past several years \citep{Ma2024-oc, Peterson-Health-Technology-Institute-AI-Taskforce2025-aq}. These models have been positioned as transformative tools capable of enhancing productivity, supporting clinical decision-making, and reducing administrative burdens. Among these applications, the generation of structured clinical notes has emerged as a popular and relatively straightforward use case. 

However, despite the technical feasibility, a number of critical challenges persist. Research has highlighted concerns about model interpretability, trustworthiness, and fit with existing workflows \citep{Farhat2024-os}. In some cases, the outputs fail to meet clinician expectations, either due to formatting mismatches, the use of inappropriate structuring (e.g., summaries instead of SOAP notes), or unexpected tone \citep{Han2024-nu}. These challenges can be further compounded by biases that are embedded in the training data used for these large-scale models, which can be particularly dangerous when applied to human behavior and groups that have historically been treated unjustly \cite{Farhat2024-os}.

A growing body of work has begun to explore not just technical performance but the human and organizational factors shaping adoption. One particularly challenging aspect is establishing healthcare provider comfort and confidence with LLM use. For example, a study by \citet{Spotnitz2024-sc} quantified the experience of medical professionals across three axes: (1) years of clinical training, (2) years of informatics training, and (3) hours of LLM use in the past 12 months. Among diverse clinical experts with novice LLM experience, the majority felt that using LLMs to support clinical tasks such as translating technical reports into layperson summaries (n=21/30) and creating discharge summaries (n=20/30) were appropriate use cases. However, tasks such as responding to patient questions about a technical report (n=7/30) or writing full reports (n=7/30) were less accepted \citep{Spotnitz2024-sc}.

This tension between the opportunity and hesitation of adoption also appears in qualitative studies. For instance, \citet{Ma2024-oc} conducted a qualitative study aiming to better understand the perceptions of mental health and AI experts when considering the integration of LLMs into the related field of mental health practice. Through semi-structured interviews, a mixture of psychiatrists (n=12), mental health nurses (n=7), and AI in medicine researchers (n=2) were engaged to understand their perspectives on integrating these tools. Participants were particularly enthusiastic about the potential for LLMs to improve efficiency and quality of work, specifically referring to how LLMs could help support the completion of routine text-generation tasks, freeing up more time and attention for patient-facing tasks. Another important finding concerned the prerequisites that should exist for integrating LLMs into the mental health field. These highlighted the necessity of relevant training sessions to prepare professionals for current and proficient use of LLMs and, in general, the enhancement of digital literacy to support their continued use despite a rapidly changing landscape. Furthermore, clinicians cited concerns around creating proper guidelines for use and management, calling for clear principles that would ensure these tools are being used responsibly \citep{Ma2024-oc}.

More specifically, some research has focused exclusively on using LLMs to enhance clinician productivity through supporting clinical note generation. For example, \citep{Han2024-nu} proposed AscleAI and conducted an interview study exploring how LLMs could support clinicians in practice. In the settings tested, challenges such as completing documentation while simultaneously engaging with patients were frequently cited. Before the introduction of LLMs, common strategies to overcome these challenges included making corrections to notes immediately following the end of a session or using specific templates to reduce time spent writing during  sessions. The outputs of the task seemed to be of particular importance, where the results were met with inconsistent feelings. Many of the concerns surrounded a mismatch of clinician expectations and model outputs. For instance, some participants were frustrated that the outputs of AscleAI were not in the SOAP format but instead provided a summary. Furthermore, some clinicians found that the inherent structuring (i.e., sentence-based vs. bullet form) did not match their expectations for how they would like their output notes to be formatted \citep{Han2024-nu}.

Given this emphasis on efficiency, it is perhaps unsurprising that the proposed use of AI-based scribes has surged in popularity. Modern AI-based scribes tout the potential to passively record patient-provider interactions and convert them into useful summarizations and documentation, simultaneously increasing patient volume while improving visit quality. Based on a recent report from the \citet{Peterson-Health-Technology-Institute-AI-Taskforce2025-aq}, approximately 60 ambient scribes are currently being applied in practice in the United States. The adoption of these scribes represents a significant deviation from the normal adoption pipeline of the industry, historically known for long implementation timelines and cumbersome workflow changes. The report suggests mixed results along different axes of improving workflows. Positive impacts have been observed for  clinical burnout, cognitive load, and the downstream quality of generated notes. Mixed feedback was observed regarding clinician experience, time saved by technology adoption, and the avoidance of ``pajama time" where clinicians complete documentation requirements outside of working hours \citep{Peterson-Health-Technology-Institute-AI-Taskforce2025-aq}.

\subsection{Nuances of Documentation in Pediatric Occupational Therapy}
Pediatric occupational therapy services are often provided in multiple distinct environments, each with unique characteristics and challenges. For instance, younger pre-school aged children are often seen through in-center and virtual consultations that are typically conducted with families present in a structured environment. Clinicians often work with families to tailor interventions to each child’s developmental needs, creating personalized treatment plans \citep{Cahill2020-bp}. In contrast, School-Based services offer in-school consultations without families present, in less structured and more rapidly changing environments. These visits can range from specific one-to-one visits or universal classroom visits where all students and educators are observed in parallel. 

Documentation in occupational therapy is a structured process of recording client information, therapeutic interventions, and treatment outcomes. At the studied center, documentation took on several forms, which depend both on the program in which the clinician (i.e. In-Center vs. School-Based) worked in and the point of treatment for the client. Progress notes are used across all visit types and are the only form of documentation legally required by the provincial OT regulatory body \citep{Amenyo2025-ku}.

Documentation in healthcare settings, particularly in occupational therapy, serves multiple critical purposes beyond simple record-keeping. The ultimate goal of this documentation is to provide comprehensive understanding of client abilities, progress, and limitations throughout their treatment journey; however, in practice this may not be the case. In its ideal form, effective documentation reflects all treatment stages from initial assessment to ongoing intervention. It captures not only physical aspects of therapy but also environmental and contextual factors affecting client participation in activities of daily living \citep{Cahill2020-bp}. Documentation acts as an essential communication tool among healthcare providers, supporting coordinated care across interdisciplinary teams. Additionally, it provides a formal legal and regulatory record that may be reviewed for accreditation purposes, used in legal proceedings, or examined for reimbursement claims \citep{Gateley2015-ap}.

While documentation remains a critical aspect of care, it has increasingly become a burden for healthcare providers at the studied center. Both clinicians and management perceived that growing documentation demands were impacting the organization’s ability to deliver services effectively. The center was facing service capacity issues, marked by long waitlists and an inability to meet the rising demand for pediatric occupational therapy services. At the same time, anecdotal evidence suggested that clinician well being is suffering, with reports of burnout and difficulty keeping up with administrative requirements. These concerns echo broader trends identified in recent reports—such as one from the Peterson Health Technology Institute AI Taskforce—which link excessive documentation to provider burnout and decreased system capacity \citep{Peterson-Health-Technology-Institute-AI-Taskforce2025-aq}.

\subsection{Case Study: Pediatric Occupational Therapy LLM Pilot Study}
\label{llmpilotstudy}
Our study was completed in collaboration with KidsAbility, a pediatric rehabilitation center that provides comprehensive rehabilitation services across several professions including occupational therapy, speech-language pathology, and physiotherapy. Within KidsAbility, Early Years (including Entry to School) and School Years programs are delivered primarily in-center, whereas the School-Based Rehabilitation program operates directly within school environments. The case study was part of a broader initiative to streamline documentation using LLMs, with the goal of enhancing service delivery capacity and reducing wait times. To evaluate this, the team conducted a pilot study using LLMs to convert therapists' brief session notes (i.e., scratch notes) into structured clinical documentation \citep{DiMaio2024-mq}. To do so the technical team adapted foundation language models which resulted in a bespoke version based on the Llama 3 8B architecture that was created purely for the process of generating visit notes from scratch notes. Model training used thousands of de-identified historical SOAP-format progress notes, supplemented with few-shot–prompted synthetic scratch notes to create paired training examples. Fine-tuning employed LoRA adapters with domain-adaptive pre-training to reduce overfitting, and the model was evaluated on a held-out test set before pilot deployment.

To evaluate both the effectiveness of this specific intervention and LLM utility more broadly ten occupational therapists participated in a comparative assessment period. In the initial phase, clinicians utilized a secure version of Microsoft's enterprise CoPilot program for three-weeks with a subsequent phase employing the custom-developed solution (three-week duration) \citep{DiMaio2024-mq}. While the Llama 3 8B model is smaller than most commercial systems, it was deliberately selected to limit overfitting during fine-tuning and to test the feasibility of a domain-specific model. For comparison, clinicians also evaluated a large general-purpose model (Microsoft’s enterprise CoPilot) under the same study conditions, attempting to distinguish size-related performance effects from broader sociotechnical factors.

\section{Methods}
It is in this context that our study was implemented. We conducted 30 interviews with 20 clinical staff members at a pediatric rehabilitation center in Ontario, Canada to learn about their perspectives on documentation and possibilities for LLMs to assist with documentation tasks. 

\subsection{Recruitment/Participants}
Participants in the study were clinical staff working for the associated organization. They were recruited by center administrative staff and referred to us for voluntary participation in the study. Participants were informed about the nature of the study and assured of confidentiality before they consented to take part in the study. In total, 20 clinicians took part in this study.

Of the 20 participants, 6 worked in the Early Years program (including Entry to School), 10 in the School-Based Rehabilitation program, and 4 in the School Years program. The Early Years and School Years programs were delivered primarily in-center, whereas the School-Based Rehabilitation program was delivered directly in school environments. Participants’ total years of practice (including experience prior to joining KidsAbility) ranged from 1 to 22 years (median = 10 years), representing a mix of junior, mid-career, and senior therapists. Ten participants took part in the pilot study (4 Early Years, 4 School-Based Rehabilitation, 2 School Years), with experience ranging from 1 to 22 years. The remaining 10 participants did not take part in the pilot (6 School-Based, 1 Early Years, and 2 School Years), with experience ranging from 3 to 15 years.

\subsection{Procedures}
Independent of this study, the center conducted a pilot workshop in which participants were introduced to two different LLMs and given instructions for how they could be used to assist with documentation tasks (see above section). Of the 20 clinicians who participated in the present study, 10 had taken part in the pilot program, and 10 had not taken part in the pilot program. Clinicians who had taken part in the pilot program were interviewed twice, once before and once after the pilot program. Clinicians who had not taken part in the pilot program were interviewed only once. 

The purpose of the semi-structured interviews was to understand the nature of the documentation burden on clinicians, their perspectives of the role of documentation in the context of their work, and their perspectives on the possible role of LLMs in helping to assist with documentation. Clinicians who had taken part in the pilot workshop were interviewed twice so that they had the opportunity to express their views before and after being exposed to two LLMs currently available to assist with documentation. In the interviews, clinicians were asked questions about their clinical practice, the kinds of documentation required for their work, what kinds of notes they take and how they take notes, and how much time documentation takes. They were also asked about whether they had previously heard about applications of AI in healthcare more generally, whether they had experience with AI or LLM tools, and whether they thought tools like ChatGPT could be appropriately used in clinical contexts. Interviews were conducted virtually and were recorded using the built in Teams functionality through Microsoft360. Interviews ranged in length from 20 to 35 minutes and were transcribed verbatim and verified by the research team.

\subsection{Analysis}
Interview transcripts were coded with the help of NVivo14.  \citet{braun2006using, braun2013successful} six-phase thematic coding process was followed to identify dominant themes in the interviews that identified the range of perspectives clinicians expressed on the issue of documentation burdens, possible ways of mitigating this burden, and the specific use of generative AI and similar technologies in this regard.

\section{Results}

Thematic analysis of our data revealed four primary themes:

\begin{enumerate}
  \item Clinical documentation in occupational therapy is highly heterogeneous
  \item The sociotechnical nature of challenges existing at the organization extend beyond the completion and creation of SOAP notes
  \item The need for flexibility and clinician autonomy
  \item The requirement of mutual learning on both the end of the clinician and the AI system to facilitate collaboration
\end{enumerate}

\subsection{Theme 1: Heterogeneity in Documentation Workflows}

Our analysis revealed substantial variation in documentation workflows across the organization, specifically when we compared conditions related to the underlying process of generating SOAP notes from scratch notes. Understanding these workflows was essential to evaluating the potential for AI-assisted documentation. Documentation workflows varied considerably from clinician to clinician but there are also notable differences in their anecdotal description of how they complete their sessions, create scratch note or memory cues, and complete their finalized documentation. Very generally speaking the documentation process typically follows this progression:
\begin{enumerate}
  \item Within Session: Creation of scratch notes or memory cues
  \item Post Session: Conversion of scratch notes into formal documentation (SOAP notes)
\end{enumerate}

Clinician accounts suggested that the execution of these tasks was influenced by both intrinsic and extrinsic factors. Intrinsic factors included prior work experience, professional training, and personal preferences for verbosity or structure. Extrinsic factors encompassed the specific program they worked within, the type of visit, client characteristics, and scheduling constraints.

A notable distinction was observed between clinicians working in the in-center programs (Early Years, including Entry to School, and School Years) and those in the School Based Rehabilitation program. One of the most significant differences involved the tools and conditions available for in-session documentation.

Clinicians in the in-center programs were able to make use of jot or scratch notes during or immediately after sessions. These notes that were created either digitally or on paper were often more substantive and detailed. However, even within this program, the format and content of scratch notes remained highly individualized.

In contrast, clinicians in the School Based program frequently conducted sessions in less controlled environments, such as classrooms with multiple children present. These conditions limited their ability to use documentation tools during the session. As a result, scratch notes -if created at all—were generally minimal, serving primarily as memory aids rather than substantive written records. This environmental constraint significantly impacted the nature and quality of intermediate documentation and could have negative downstream impacts in a tool that was explicitly designed to map scratch notes to completed SOAP notes. Furthermore, in the School-Based program clinicians employ a tiered approach to care, where visits often consist of observing classroom settings rather than in a one-to-one session.

This contrast in session context and documentation opportunity is captured in the following reflection from a School-Based occupational therapist:

\begin{quote}
\textit{``So I work almost 100\% in schools. I don't do a lot of like one to one treatment, so I wouldn't have a kid for like a session like a clinician in center might, but I would either be embedded in the classroom and then like have a space to work ideally or I would like a rare occasions would pull a kid into like a small space to do some work with, but for the most part I'm in their classroom context or in like a separate spot on the school working on something with that kid.''
-P08}
\end{quote}

The less predictable nature of School-Based work also shaped when documentation could occur. Early Years clinicians often had scheduled time between appointments to write notes. In contrast, School-Based clinicians faced variable schedules and frequent coordination with school staff.

Continuing this point, Participant 08 explained how scheduling constraints affect documentation timing:

\begin{quote}
\textit{``Some schools will have me booked in like a classroom, but for the most part I sort of schedule it on the fly so like I keep a running tally of things that I need to do in my head. Go and check in on those kids check in on their staff. See if there's anything new, any like priorities and then do that sort of for several hours within the morning, and then by the time we hit about like 1:30-2 o'clock, their break times are they're sort of winding down the day. And I have to start just doing all the paperwork associated with that day.''
-P08}
\end{quote}

The main findings from this theme suggest that workflows and intermediate documentation artifacts are not only highly personalized but are also deeply shaped by programmatic context. These variations include differences in scratch note structure, the timing of documentation activities, and the use of shortcuts or templates to streamline note conversion. The combination of these factors introduces a significant challenge for general-purpose or even bespoke LLM systems, as inconsistency in both input (scratch notes) and expected output (SOAP notes) reduces the likelihood of generating consistently high-quality results.

\subsection{Theme 2: ``SOAP notes aren't the problem''}

Many participants demonstrated a clear awareness of the broader challenges facing the organization. Our analysis revealed that the core issues clinicians face are not rooted in the act of writing SOAP notes, but stem from a broader set of systemic and contextual factors. Some of the key factors include the heterogeneity of clinicians, organizational policies and decisions, as well as technology factors.

One in-clinic therapist framed the issue plainly, shifting the focus away from writing toward broader workload challenges:

\begin{quote}
\textit{``No, I think if I had to tell you what I think the problem is, I don't think it's our ability to write, It's the amount of things we have to do.''
-P03}
\end{quote}

This sentiment was echoed by another participant (P05), who initially approached the LLM documentation tool with optimism but later found it failed to address what they perceived to be the real bottleneck:

\begin{quote}
\textit{``I think I was intrigued by it at first and then the reality of it, I was like, it actually doesn't save me time, It didn't save me time, in terms of the soap notes, anyways from that was my experience with it that it was I think our other processes are extremely inefficient and you know but I think doing the soap note isn't necessarily the thing that takes the time''
-P05}
\end{quote}

Although many participants described a challenging relationship with documentation, they also expressed a sense of pride, ownership, and value in the notes they created. Several emphasized that writing SOAP notes is a professional obligation governed by the College of Occupational Therapists and something they take seriously and value with ongoing visits.

For instance, P01 highlighted how notes served as a crucial memory aid in ongoing care:

\begin{quote}
\textit{``It's definitely a burden in some sense, because it does take quite a bit of time to get those notes done, but I find the notes that I write as sole valuable when I go see that family again. So when I'm seeing like 60 families trying to remember who you did what with and like most of our kids are two, three years of age.''
-P01}
\end{quote}

Another clinician reflected on the personal voice embedded in documentation, expressing concern about AI-generated text lacking their distinctive tone:

\begin{quote}
\textit{``Like even in an objective way where I'm stating objective, like you know I'm not being biased. Well, I'm biased inherently, but you're trying to be objective, you're showing measure like you're reporting. It still has your person like, it has your sound to it. And so when you're reading and it doesn't have your sound, like this is wild.''
-P08}
\end{quote}

A consistent thread across interviews was a concern with how time is structured and constrained, rather than with documentation itself. Many felt they lacked sufficient time to effectively balance the direct and indirect tasks required to deliver high-quality care. Several clinicians, particularly those in the School-Based program, noted that organizational policies limited their ability to create notes that added value beyond regulatory requirements. For example, P08 described how institutional expectations around formatting and dissemination reduced their motivation to engage deeply with the documentation:

\begin{quote}
\textit{``...if I've done my documentation that day and kept up on it, I have to do it again in these visit notes and our process to disseminate them is annoying. So not only is writing them annoying, but getting them out is annoying or has historically been annoying and therefore I don't like to do them. But the other thing that like because of the way these are structured and the way they are sent out, it's just as fast for me to write an email and that's how parents wanna read it.''
-P08}
\end{quote}

One factor contributing to clinicians' feelings of overwhelm and burnout appears to be the recent transition to a new health record platform and the complexities associated with documenting within it. While intended to streamline workflows, many clinicians found the system clunky and inefficient—particularly when completing notes that required duplicating information across multiple forms. For example, many clinicians noted that completing notes often involves repetitive clicking to input each component of a SOAP note, which they found both frustrating and time-consuming. One clinician described this repetitive structure as a fundamental barrier to efficiency:

\begin{quote}
\textit{``So you've got the pre-assessment form, you've got a soap note and you've got a consent form that are all documenting the same interaction.''
-N03}
\end{quote}

Another clinician reflected on how the platform's inefficiencies compounded existing frustrations but reinforced that writing notes, in itself, wasn’t the issue:

\begin{quote}
\textit{``Yeah, I think, I mean, I think it further like confirms that [EHR SOFTWARE] is extremely inefficient, like there's inefficiencies built into a lot of our processes, which are getting in the way of us being efficient, but actually our ability to write the notes isn't necessarily [THE PROBLEM] Right?''
-P05}
\end{quote}

Together, these reflections complicate the assumption that LLMs simply need to automate SOAP notes to be useful. As participants emphasized, the real barriers to efficiency lie in systems, structures, and workflows that shape how, when, and why documentation gets done.

\subsection{Theme 3: Flexibility of Proposed Tools and Promoting Clinician Autonomy}

A strong theme that emerged was clinicians’ dedication to their work and the sense of professional identity they brought to how they performed it. Many participants had years of experience and had developed highly individualized workflows and informal tools to manage both direct and indirect aspects of care. Among these were customized documentation templates and strategies that they had refined over time. For example, several clinicians described how they maintained external resources—such as written prompts or structured text banks—outside of the official clinical record to make documentation more efficient and personally meaningful. As one participant explained:

\begin{quote}
\textit{``I feel like I mean some of it is changing with the forms, but we've all had kind of our own little templates of things that we would use to support our documentation.''
-N01}
\end{quote}

Another clinician elaborated on the cognitive benefits of these personalized prompts:

\begin{quote}
\textit{``I've always lived with templates outside of the clinical record because I found that they have been the most efficient for me. They helped to prompt my memory of what I need to ask, what I need to say, what I need to write''
-N02}
\end{quote}

These reflections echo themes from earlier in the analysis, particularly the heterogeneity of documentation workflows described in Theme 1. Yet here, clinicians emphasized the value of autonomy in shaping these workflows, adapting them in real time to meet client needs. For example, even if a clinician typically relies on a specific tool during a session, they may choose to forgo it if they believe it could negatively impact the client's experience. In such cases, they may conclude sessions without the usual steps that support their documentation workflow. 

This adaptive flexibility also shaped clinicians’ perceptions of the LLM documentation tool piloted in the study. While many began the trial with high hopes, several felt that the tool did not yet align with their needs or standards. One Early Years clinician reflected on their disappointment:

\begin{quote}
\textit{``Honestly, for me, I was hoping for a better outcome like for me was a bit disappointing. I was really wanting it to be helpful and a little bit more efficient for me and then I just, I don't know if I think we'll get there for me it's not ready, It's not ready for where I think it needs to be, so for me it was.''
-P01}
\end{quote}

To thrive in this environment, clinicians need the autonomy to determine when and how to use tools that support their documentation practices. During the pilot, several clinicians expressed that there were situations in which the use of LLMs did not feel appropriate for the specific context, yet they felt pressured to integrate them into their workflow. As the pilot progressed,  some participants began selectively disengaging from the tool’s structured workflow. One therapist (P02) described situations where they intentionally bypassed the LLM system, especially when feeling overwhelmed:

\begin{quote}
\textit{``Yes, I would say a lot of the time with [the LLM tool], I ended up doing that maybe like 50 to 75\% of the time I was doing it for myself because just the when the notes came through, if I was, like flustered with something else I was like, hey, I'm just gonna do this on my own because, like, I don't have the energy to, like, go through and fix this note again.''
-P02}
\end{quote}

Rather than signaling disinterest in AI altogether, these decisions often reflected a desire for more control. Several clinicians expressed curiosity about how generative AI might be integrated more usefully into their broader workflows—particularly when they had flexibility in how the technology was used.

This openness to experimentation extended beyond SOAP notes. Clinicians who were less invested in the pilot’s primary goal still imagined other ways LLMs could support their work, including automating family-facing documentation or helping create personalized resources for clients.

As one clinician summarized:
\begin{quote}
\textit{``I know you're focused on documentation, but I see some huge advantages for helping clients to overcome barriers, be more successful in you know their life and their goals.''
-N01}
\end{quote}

These reflections reinforce the notion that tool usability alone is not enough. Meaningful adoption of LLMs in clinical practice will depend on whether the systems promote, and not constrain, clinician autonomy and align with the individualized, context-sensitive nature of care.

\subsection{Theme 4: Learning on Both Sides}

As discussed in Theme 3, clinicians showed a strong interest in exploring potential applications of LLMs beyond the prescriptive documentation use case. While the pilot focused on converting scratch notes into SOAP notes, several participants experimented with broader applications, testing the limits of the tool and imagining alternative integrations (see Key Point 1, Table 5). These clinicians expressed excitement about the long-term potential of LLMs—particularly if they could be shaped to fit their individualized workflows. For example, one clinician envisioned a tool that could not only transcribe input but intelligently organize it into clinical formats:

\begin{quote}
\textit{``It's the repetitive process, like if I had a smart notebook that I could write some things down and it puts it into typed like that would be great. But also like, yeah, if it could somehow integrate that into my template like ohh it knows that this is like a subjective thing or it knows that this thing is goes into the objective category, this part goes into analysis like how cool would that be?''
-N02}
\end{quote}

Despite this curiosity, the pilot’s input method, where the clinicians were asked to input raw scratch notes, posed barriers. Scratch note practices varied significantly across participants: some used full sentences, others employed shorthand or symbols, and some relied entirely on memory. This variability made standardized inputs difficult, leading to mixed results. As expected, clinicians whose habits already aligned with the input assumptions of the tool adapted more easily than those whose workflows diverged.

In order to participate, some clinicians, particularly those who typically relied on sparse or no scratch notes, created a new intermediary artifact: refined notes written solely for use with the LLM. This additional step disrupted their established routines and introduced inefficiencies. When asked about the importance of scratch notes, participants consistently emphasized that scratch notes serve a singular, simple purpose: to support accurate post-session documentation. These notes are typically informal, not saved, and rarely revisited. Formalizing them as part of the pilot process fundamentally altered how they were produced, leading to frustration. Rather than change their in-session practices, many clinicians chose to create an extra, more polished version of their notes specifically for the system.

\begin{quote}
\textit{``I'm like this is too much effort for me to tell you what to do. I already know what I wanna do, so I would just use my own note like I would just abandon it altogether because it was like this is taking me too long in terms of what I want this note to look like.''
-P01}
\end{quote}

Another participant echoed this, explaining that the cognitive load of preparing model-friendly inputs nullified any benefit:

\begin{quote}
\textit{``I would just end up like scribbling notes. So, then to take that information and type it, it was sort of like a may as well, if I'm typing it anyways, I may as well just type it into SOAP format, right? So and then or if I do a phone call that I do tend to type, but as I type, I kind of organize it like sort of anyways. So I think, yeah, I found it didn't actually really save me time and I ended up editing everything to go back to my usual format.''
-P05}
\end{quote}

Many clinicians expressed skepticism that the model could handle their typical scratch notes, leading them to write much more detailed input than usual. In many cases, rather than submitting authentic scratch notes, clinicians preemptively wrote content that resembled finalized SOAP notes and used those as inputs. Ironically, this strategy transformed the LLM into a formatting assistant rather than a generative tool and often eliminating any time savings. This shift was especially apparent when participants were testing the bespoke model trained on historical documentation pairs, where clinicians feared minimal input might lead to hallucinations or clinical inaccuracies:

\begin{quote}
\textit{``Sometimes I put it into like SOAP format a little bit... but sometimes it was kind of in sections but it just didn't have the headings, which probably confused it. But yeah, I think I varied—like sometimes I would sort of do it as a SOAP, put it in and say like, see what it gives me. But then things ended up getting kind of shifted around and so I would kind of revert back to what I did before.''
—P05}
\end{quote}

This lack of trust reflects a deeper misalignment between clinician expectations and system capabilities. Many participants expressed discomfort with the idea that an AI system could or should contribute substantively to clinical documentation. This was rooted in concerns about professional responsibility, legal risk, and patient trust:

\begin{quote}
\textit{``No, I mean like parents come to come to us because they like you're the expert, you know what we're doing? And I don't know how they would feel if I said 'Yeah, some computer model generated these recommendations'.''
-P01}
\end{quote}

The AI systems also struggled when inputs deviated from what they had been trained or tuned to expect. Several clinicians reported that models miscategorized content or omitted key information:

\begin{quote}
\textit{``Like it just kind of would put stuff in areas in the headings that it didn't belong. So then I would have to go through and edit it and then it would put nothing in my subjective even though a lot of my notes are like heavily subjective because it's like updates from family and then it would put it all in like the plan which isn't where it's supposed to go.''
-P02}
\end{quote}

\begin{quote}
\textit{``I found that it just caused a lot of errors in writing. So, like sometimes if you said like the child was anxious, it would change it to the child has anxiety and like that's a big known like I can't say that.''
-P03}
\end{quote}

\begin{quote}
\textit{``I found it mixed things up a lot like it was only subjective report and then it was saying that I observed certain things or sometimes they're just random information in there that was not I'm like relevant or I found sometimes you know, like I put OT recommendations and listed things I recommend and those were like, gone with both models.''
}-P04
\end{quote}

Beyond the technical issues, participants raised concerns about the burden of adopting new tools in already demanding clinical environments. Several clinicians noted that they simply lacked the time or cognitive bandwidth to meaningfully engage with the pilot while maintaining their usual responsibilities. Despite these challenges, some clinicians remained open to the idea of AI as a thought partner—not just a documentation assistant. 

As one clinician speculated:
\begin{quote}
\textit{``And I was like, this is really interesting. I remember saying to him at that time, like in some ways, like I think it could skew your information, but I also feel like—I wonder if we could learn from it. In that, like, when they're doing an analysis and stuff like that, is it going to think of things that I wouldn't have thought of?''
—N02}
\end{quote}

This comment illustrates a subtle but important shift: a move from viewing AI as a documentation tool toward viewing it as a potential clinical collaborator.

\section{Discussion}
Our findings show that clinician responses and experiences with AI-assisted documentation cannot be fully explained by the technical capacity of the systems alone. Instead, these responses are shaped by a web of social, organizational, and technical factors, as well as the fit between them. To interpret our results, we draw on the FITT framework \citep{Ammenwerth2006-ap}, which conceptualizes technology adoption as a function of the alignment—or misalignment—between individual users, the tasks they perform, and the technology they are asked to adopt. This framework positions sociotechnical fit as a continuum, highlighting the interactions between each of the three objects.

We organize this discussion using the FITT framework, with one section each for Individuals and Organizational Context, Task Complexity and Variation, and Technology Design and Assumptions. We conclude with an analysis of how misalignments across these elements produced integration challenges and what this implies for the deployment of LLMs in healthcare.

\subsection{Individuals and Organizational Context (Individual Dimension)}
From a social lens, it is essential to recognize that users of clinical documentation systems operate within diverse personal and professional contexts. Each clinician brings a unique combination of preferences, constraints, and experiences to their workflow. These differences may stem from variations in clinical training environments, years of professional experience, personal stressors, and the specific demands of their practice settings.

Our analysis underscored the heterogeneity in documentation workflows among clinicians. Importantly, we found no single optimal approach to completing documentation tasks. Instead, workflows were highly individualized and often adapted over time. As such, it is counterproductive to mandate a uniform approach to intermediary tools—such as scratch notes—used during documentation. Rather, system design should remain flexible and responsive, enabling clinicians to articulate and shape the ways in which AI systems can support their practices.

Almost all clinicians we spoke to emphasized their legal obligation to maintain good record keeping through documentation; however, this meant different things to different individuals. Furthermore, different clinicians have different comfort levels giving up control of their college mandated documentation. Some clinicians felt that their clinical identity was reflected in their documentation and for others this was not the case.

Clinicians' approaches to documentation are shaped in part by the training they received throughout their careers, which varies considerably across individuals. Some clinicians were explicitly trained to use intermediary tools such as scratch notes to support documentation whereas others were not, and instead only do what is necessary to create the final documentation. Within our case study, internal mentorship also played a role in shaping documentation practices. Senior clinicians often pass down tools, strategies, and templates to newer staff, contributing to informal but influential norms around documentation.

Organizational policies were also highlighted as a major barrier to successful participation and uptake of the proposed AI systems. Two features were prominently highlighted here: (1) the number of documentation artefacts that is required and (2) the recent transition to the new electronic health record platform. Both factors contribute to a broader perception among clinicians that current documentation practices are inefficient and burdensome. Participants expressed that these challenges could be improved through review of organization policy that minimizes the number of tasks required and reanalysis of workflows for documentation in the prescribed platform. These concerns were particularly pronounced among clinicians in School-Based Programs, where the requirement to generate multiple documentation artefacts was commonly cited.

Additional barriers were highlighted for the School-Based program. In our interviews clinicians emphasized the extremely dynamic and unpredictable environment in which they deliver therapy. They suggested that this makes it difficult to complete documentation at regular or scheduled intervals. Many clinicians reported that they had insufficient time to complete documentation due to constraints imposed by the daily variability of their settings—for instance, navigating the expectations of different schools, teachers, or institutional routines.

\subsection{Task Complexity and Variation (Task Dimension)}

Although documentation is often described as a routine or low-risk area for incorporating AI, our findings show that it is far from standardized in practice. The nature of clinical documentation varies significantly depending on the setting, program, clinician role, and even client type. Within the FITT framework, such variation reflects the complexity of the task itself.

In this study, the task of transforming scratch notes into SOAP notes was assumed to be consistent across contexts. However, this assumption did not hold. In the Early Years program, clinicians had access to tools and time to create more structured scratch notes during sessions. In contrast, clinicians in School-Based programs often delivered care in unpredictable classroom environments and rarely produced structured notes until much later, if at all.

Scratch notes were also not uniformly understood across the organization. Some clinicians used shorthand memory aids, while others relied on mental recall. There was no shared standard for what constitutes a high quality scratch note or how it should be used. This heterogeneity in intermediary artifacts directly impaired the perceived relevance and usefulness of LLM-based tools.

\subsection{Technology Design and Embedded Assumptions (Technology Dimension)}

The third FITT component—technology—considers how well a system’s capabilities and constraints align with user expectations and task requirements \citep{Ammenwerth2006-ap}. In this case, many of the technical challenges presented by participants arose from pilot design assumptions that did not align with their actual workflows. This is a commonly documented phenomenon in AI systems known as the translation gap, where systems that perform well in highly controlled, benchtop settings fail to deliver similar results with real users upon deployment. When developing an AI system, the practitioner embeds inductive biases into the system through the modelling decisions they make. In the case of the LLM pilot program, several assumptions may have contributed to the relatively low clinician enthusiasm for using the tools.

The ability of AI systems to work within a wide variety of environments is known as the model's capacity to generalize. The structure of the pilot implicitly assumed that input and output formats were sufficiently consistent across the organization and among clinicians. However, our analysis highlighted significant heterogeneity in both documentation practices and clinical roles at the partner organization, suggesting that a single AI system is unlikely to support documentation workflows organization-wide. In particular, the program in which a clinician works has a significant impact on both the input (i.e., scratch note) and the required output documentation (e.g., SOAP note, visit note). This variation makes it difficult to design a model that performs well across all scenarios—especially in the absence of additional context indicating that scratch notes may be sparse or formatted differently than the examples used during training.

One of the most significant inductive biases embedded during AI system training is the selection of specific training data. In the case of the bespoke LLM fine-tuned on historical scratch–SOAP pairs, an implicit assumption about what constitutes a ``good'' scratch input and a corresponding SOAP output was reinforced. Our analysis suggests that this may have posed challenges for clinicians, as there are no explicit organizational standards for what a scratch note should include—or whether such notes should be created at all. Additionally, clinicians held varying perspectives on what makes a high-quality SOAP note. When either of these expectations diverged from the examples provided during fine-tuning, the model often failed to perform as intended.

For the general-purpose model, Copilot, it is important to recognize that these systems were not trained with clinical documentation in mind. They were developed using general-purpose internet text, which means that while fewer inductive biases were embedded, the models also offered less guidance on what was or was not appropriate. The only context available to the model came through direct prompting.

Both models operated under the assumption that clinicians would provide an intermediary document in the form of a scratch note containing meaningful information to support LLM-generated documentation. However, when this assumption did not hold, many clinicians introduced an additional step of reformatting or rewriting their scratch notes to better align with what they believed the model required. In some cases, this involved providing excessive detail and effectively completing the SOAP generation process manually before even inputting the note into the model.

\subsection{Misalignment and Integration Failures}
The full picture of adoption became more clear when we considered how these three dimensions interact, not how any single one behaves in isolation. The FITT framework emphasizes that integration depends on the degree of fit between individuals, tasks, and technology \citep{Ammenwerth2006-ap}. In our study, misalignment at multiple intersections led to breakdowns in adoption and integration.

First, there was poor \textit{Technology–Task fit}: the tools assumed structured scratch notes, but clinicians’ inputs were often sparse, informal, or nonexistent. Second, there was weak \textit{Technology–Individual fit}: some clinicians felt the tool undermined their control over professional documentation, or demanded extra work without sufficient benefit. Finally, there were strains in \textit{Individual–Task fit}, as clinicians had varied ideas about what good documentation meant and how much effort it warranted.

These misalignments created friction that revealed a lack of true workflow integration. Clinicians adapted by reformatting notes or supplying overly detailed inputs, often in an effort to improve system performance. This overcompensation reflected a trust calibration issue: clinicians overestimated the AI’s capabilities and adjusted their behavior accordingly, but the system wasn’t designed to handle such inputs. As a result, rather than reducing effort, the tools often introduced additional work and complexity.

A critical sociotechnical issue was the value misalignment between what clinicians wanted from AI assistance and what the tools provided. Clinicians varied in their comfort levels with AI supported documentation. This disconnect shows how the technology addressed aspects of documentation that clinicians valued as part of their professional identity rather than automating the truly burdensome administrative tasks.

The concept of change fatigue was also apparent in our analysis, as the AI implementation occurred shortly after the transition to the the new electronic health record platform. This timing created additional resistance, with clinicians feeling overwhelmed by consecutive technological changes without sufficient adaptation time. This illustrates how organizational context significantly influences technology adoption regardless of the tool's technical capabilities.

We observed a mismatch in adaptation between users and the technology. Clinicians, in an effort to support system performance, often introduced an additional step of creating more structured or ``refined'' scratch notes specifically for the AI tools. While this demonstrated clinicians' willingness to adapt, the AI systems were not designed to handle the diversity of these modified formats.

\section{Conclusion}

This study set out to examine whether LLMs could assist clinicians in one of the most routine and structured documentation tasks in healthcare: generating SOAP notes. Despite the narrow scope and low-risk nature of this pilot, our findings uncovered persistent sociotechnical barriers that complicate the integration of generative AI into clinical workflows.

Drawing on qualitative analysis and the FITT framework \citep{Ammenwerth2006-ap}, we observed that successful adoption hinged not solely on technical performance, but on the complex interplay between individuals, tasks, and technology. Clinicians varied in their professional training, clinical contexts, and expectations of what documentation should achieve. The task itself, often framed as standard, was in practice deeply contextual, shaped by organizational demands, time pressures, and personal documentation habits. In contrast, the LLM tools operated on idealized assumptions about user behavior, note structure, and workflow uniformity—assumptions that failed to hold in the real-world environment.

What emerged was a portrait of clinical documentation as a sociotechnical task. The heterogeneity of clinical workflows, varying documentation requirements across programs, and distinct individual preferences all contributed to an environment where a one-size-fits-all AI solution is unable to succeed. This analysis highlights the challenges clinicians face extend beyond the documentation process itself to include organizational policies, workflow inefficiencies, and technology integration issues that cannot be solved by AI tools in isolation. AI tools that fail to account for this complexity risk not only low adoption but also erosion of professional trust. 

To move forward, we argue for a more holistic strategy to deploying and developing LLMs for clinical workflows and documentation. These deployment strategies must balance technological innovation with flexibility, clinician autonomy, and organizational readiness. This includes building adaptable tools that can respond to diverse workflows, investing in clinician-facing AI literacy, and fostering ongoing collaboration among clinicians, developers, and administrators.

We present the case of pediatric occupational therapy documentation not as an isolated incident, but as an early warning. If we are still struggling to integrate LLMs into something as well-scoped and “simple” as note-generation, we must question its readiness for more complex and higher-risk applications. This study highlights that successful integration depends not just on what current models are able to do, but on how well it aligns with the realities of clinical work. Responsible deployment in healthcare needs to start with a better understanding of the people, practices, and systems these tools are meant to support.

\section{Acknowledgments} 
This research was supported by KidsAbility and Mitacs. We thank Brendan Wylie-Toal, Ilona Koshy, Nikita Gaikwad, the KidsAbility Innovation Team, and the participating clinicians for their collaboration and valuable insights.

\bibliography{AIES}

\end{document}